\documentclass[a4paper,aps,pre,superscriptaddress,floatfix,nofootinbib,twocolumn,bibtex]{revtex4}

\usepackage{bbold}
\usepackage{bbm}
\usepackage[pdftex]{graphicx}
\usepackage{latexsym,amsmath,verbatim}
\usepackage{color}
\usepackage{rotating}
\usepackage{verbatim}
\usepackage{multirow}
\usepackage[english]{babel}
\usepackage{comment}

\begin{document}

\title{Percolation in real interdependent networks}

\author{Filippo Radicchi}
\affiliation{Center for Complex Networks and Systems Research, School of Informatics and Computing, Indiana University, Bloomington, USA}
\email{filiradi@indiana.edu.}


\begin{abstract}
\noindent
The function of a real network depends not
only on the reliability of its own components, but
is affected also by the simultaneous operation
of other real networks coupled
with it. 
Robustness of systems composed of
interdependent network layers 
has been extensively studied in recent years.
However, the theoretical frameworks developed so far
apply only to special models in the
limit of infinite sizes. 
These methods
are therefore of little help
in practical contexts, given that real interconnected
networks have finite size and their structures are
generally not compatible with those of graph
toy models. 
Here, we introduce a theoretical method that takes as
inputs the adjacency matrices of the
layers to draw the entire phase
diagram for the interconnected network,
without the need of actually simulating
any percolation process. We demonstrate that
percolation transitions in arbitrary interdependent networks
can be understood by decomposing these system
into uncoupled graphs: the intersection among the layers, and
the remainders of the layers. When the intersection dominates the
remainders, an interconnected network undergoes
a continuous percolation transition.
Conversely, if the intersection is dominated by the contribution
of the remainders, the transition becomes
abrupt even in systems of finite size. 
We provide examples of real systems
that have developed interdependent networks
sharing a core of ``high quality'' edges 
to prevent catastrophic failures.
\end{abstract}

\maketitle

Percolation is among the most studied topics in
statistical physics~\cite{stauffer1991introduction}.
The model used to mimic percolation processes
assumes the existence of an underlying
network of arbitrary structure.
Regular grids are traditionally considered to model
percolation in
materials~\cite{kirkpatrick1973percolation, berkowitz1995analysis}.
Complex graphs are instead
assumed as underlying supports
in the analysis of spreading phenomena
in social environments~\cite{pastor2001epidemic, newman2002spread},
or in robustness studies
of technological and infrastructural
systems~\cite{albert2000error, cohen2000resilience, callaway2000network}.
Once the network has been specified,
a configuration of the percolation
model is generated assuming 
nodes (or sites) present with probability $p$.
For $p=0$, only a disconnected configuration
is possible. For $p=1$ instead,
all nodes are within the same
connected cluster. As the occupation probability varies,
the network undergoes a
structural transition  between
these two extreme configurations.
Although there are special substrates, e.g., one-dimensional
lattices, where the percolation transition may be
discontinuous, in the majority of the cases,
random percolation models give rise to
continuous structural changes~\cite{dorogovtsev2008critical}.
This means that the size of the largest
cluster in the network, used as a
proxy for the
connectedness of the system, increases from the
non-percolating to the percolating phases
in a smooth fashion.

\begin{figure}
\begin{center}  
\includegraphics[width=0.45\textwidth]{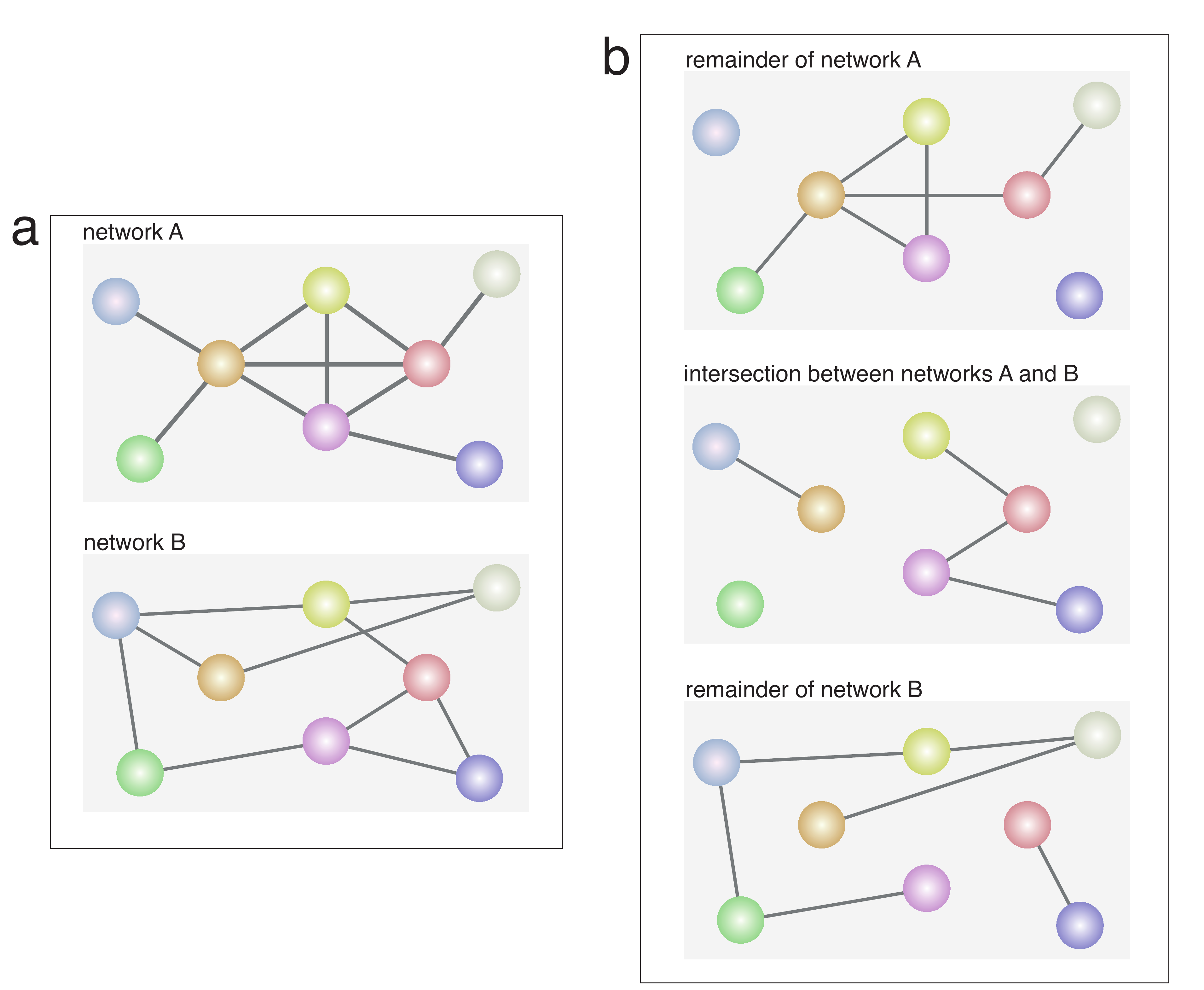}                                                                                 
\end{center}                                                                                                                     
  \caption{Decomposition of interconnected networks
    into uncoupled graphs.
{\bf a)} Schematic example of two coupled networks 
$A$ and $B$. In this representation, nodes 
of the same color are one-to-one interdependent. {\bf b)}
In the percolation model, 
the interconnected system is equivalent
to a set of three
graphs that do not share any edge: the remainders of
the network layers $A$ and $B$, and their intersection.
}
\label{fig:1}
\end{figure}

The percolation transition
may become discontinuous in a slightly
different model involving not just a single
network, but a system composed of two or more
interdependent graphs~\cite{buldyrev2010catastrophic}.
This is a very realistic
scenario considering that
many, if not all, real graphs are ``coupled''
with other real networks. Examples can be found in several
domains: social networks (e.g., Facebook, Twitter, etc.)
are coupled because they share the same
actors~\cite{szell2010multirelational}; multimodal transportation
networks are composed of different layers (e.g., bus, subway, etc.)
that share the same
locations~\cite{barthelemy2011spatial,de2014navigability};
the functioning of communication and power grid systems
depend one on the other~\cite{buldyrev2010catastrophic}.
In the simplest case, 
one considers an interconnected
system  composed of only two network
layers. Nodes in both layers are uniquely
identified, but the
way they are connected to the other vertices
is not necessarily identical (see Fig.~1).
In the percolation model defined on this system,
nodes are still present with probability $p$.
Since the networks are interdependent,
the presence of a node in one layer
implies the presence of the same vertex in the other
layer. However, as $p$ varies, one node may not be
simultaneously within the largest clusters
of both layers. In such a case, the
vertex is said to be outside the
largest cluster of mutually connected nodes.
This is a set of nodes identified in a recursive manner,
and composed of vertices that are simultaneously in the
largest clusters of both network layers thanks
only to connections with other nodes within the set.
It has been proved that, in infinitely large interconnected systems
composed of two uncorrelated random networks, the
percolation transition, monitored through the
size of the largest cluster of mutually connected
nodes, is discontinuous~\cite{buldyrev2010catastrophic,
  gao2012networks, son2012percolation}. This
result has been however shown to not apply
to  more general network models that account
for degree correlations~\cite{PhysRevX.4.021014, reis2014avoiding}.
Unfortunately, all these theoretical approaches have been
developed under two special, and unrealistic,
assumptions. First, they hypothesize that
network layers are generated according to
some kind of graph toy model whose topology
is not specified by a one-zero adjacency matrix,
but rather a list of probabilities for pairs of
nodes to be connected. Second, they apply
only to the case of infinitely large systems.
Real interdependent networks, on the other hand,
are composed of layers very different
from those that can be generated with
toy models, and they clearly have finite size.
In this paper, we
introduce a novel theoretical approach
of direct applicability to the study
of percolation transitions in
real interdependent networks.

To illustrate our methodology, we start 
from a
simplified version of 
a recent theory
developed for percolation
in real 
isolated
networks~\cite{bollobas2010percolation, PhysRevLett.113.208702,
PhysRevLett.113.208701}.
Consider an undirected and unweighted
graph 
composed of $N$ nodes and $E$ edges.
The  structure of the graph is
encoded by the adjacency matrix $A$,
a symmetric $N \times N$ matrix whose
generic element $A_{ij}$ differs
from zero and equals one only if
vertices $i$ and $j$ share an edge.
Without loss of generality, we assume
that, when all nodes are present in the network,
the graph is formed by a
single connected component.
Let us consider an arbitrary value
of the site occupation probability $p \in (0,1)$,
and indicate with $s_i$ the probability that
the generic node $i$ is part of the largest
cluster. The order parameter
of the percolation transition is simply defined as the average
of these probabilities over all nodes in the graph, i.e., 
$P_\infty = \frac{1}{N} \sum_i s_i$.
Note that $s_i$ is a function of $p$, but, in the following,
we omit this dependence for shortness of notation.
As a first attempt, we can say that
the probability $s_i$ for node $i$ to be part of
the largest cluster is given by
\begin{equation}
  s_i = p  \, [ 1 - \prod_{j \in \mathcal{N}_i} \, (1 - s_j) \, ] \; ,
  \label{eq:site_dense}
\end{equation}
where $\mathcal{N}_i$ is the set of neighbors
of vertex $i$. 
The probability $s_i$
is written as the product of two contributions:
(i) the probability that the node
is occupied; (ii) the probability that at least one
of its neighbors is part of the largest cluster.
The attempt of Eq.~(\ref{eq:site_dense_vec})
relies on the so-called locally tree-like
approximation~\cite{dorogovtsev2008critical}. In this ansatz, 
neighbors of node $i$ are not directly connected, and
this allows us to consider the probabilities $s_j$
as independent variables. This approximation
typically holds in real
networks~\cite{PhysRevLett.113.208702}, but
does not apply to regular lattices.
Introducing the vectors $\vec{u}$ and $\vec{q}$, whose $i$-th
components are respectively $u_i = \ln(1-s_i)$ and
$q_i = \ln(1 - s_i/p)$, 
we can write the set of coupled equations~(\ref{eq:site_dense})
into the single vectorial equation
\begin{equation}
  \vec{q} = A \, \vec{u} \; 
  \label{eq:site_dense_vec}
\end{equation}
A trivial solution of Eq.~(\ref{eq:site_dense_vec})
is given by the configuration
$\vec{u} = \vec{q} = \vec{0}$, corresponding
to $\vec{s} = \vec{0}$ or
$s_i=0$ for all $i=1, \ldots, N$. In the
proximity of this configuration, we can
make use of the truncated Taylor expansion
$\ln{(1-x)} = - x$, and
Eq.~(\ref{eq:site_dense_vec}) reads as
\begin{equation}
  \vec{s} = p  A \vec{s} \;,
  \label{eq:stability_site_dense}
\end{equation}
thus an eigenvalue/eigenvector equation.
By the Perron-Frobenius theorem, 
the only solution having a physical
meaning of this equation is obtained
by setting $p = 1/\lambda$ and $\vec{s} = \vec{l}$,
with $(\lambda, \vec{l})$ principal eigenpair
of the adjacency matrix $A$. This tells us
that the solution of Eq.~(\ref{eq:site_dense})
is $s_i = 0$, for all $i=1, \ldots, N$,
if the site occupation probability is
smaller than $1/\lambda$. In this region,
the network is in the non-percolating regime.
Slightly on the right of
$1/\lambda$, the vector
of probabilities $\vec{s}$ starts
to grow in the direction of the principal
eigenvector of the adjacency matrix, and
the order parameter is not longer zero.
For any value of the site occupation probability
larger than $1/\lambda$, the network is in the
percolating phase. The site percolation
threshold obtained using the approximation
of Eq.~(\ref{eq:site_dense}) is thus $p_c = 1/\lambda$.
Further, Eq.~(\ref{eq:site_dense}) can be solved
numerically to draw the percolation diagram
of the network in this approximation.

The most serious limitation
of Eq.~(\ref{eq:site_dense}) is
to introduce a positive feedback among
probabilities. An increment in the probability
$s_i$ produces an increase in the
probabilities $s_j$ of the neighbors,
which in turn causes an increment in the
probability $s_i$, and so on. To avoid the presence
of this self-reinforcement effect, we
can rewrite Eq.~(\ref{eq:site_dense}) as
\begin{equation}
  s_i = p \,[\, 1 -  \prod_{j \in \mathcal{N}_i} \, ( 1 - r_{i \to j}) \, ] \; .  
  \label{eq:site_sparse1}
\end{equation}
This equation still relies on the
locally tree-like ansatz.
Here, $r_{i \to j}$ stands for the probability that node $j$
is part of the largest cluster independently
of vertex $i$. We note that, while this quantity can
be defined for any pair
of nodes, 
only contributions given by
adjacent vertices 
play a role in Eq.~(\ref{eq:site_sparse1}).
We can think $r_{i \to j}$ as one of the
$2E$ components  of a vector $\vec{r}$.
In the definition of $\vec{r}$, 
every edge $(i,j)$ in the graph is
responsible for two entries, i.e., $r_{i \to j}$ and $r_{j \to i}$.
For consistency, the
probability $r_{i \to j}$ obeys
\begin{equation}
  r_{i \to j} = p \, [ 1 - \prod_{k \in \mathcal{N}_j \setminus \{i\}} \, ( 1 -  r_{j \to k}) \, ] \; .
  \label{eq:site_sparse2}
\end{equation}
The product of the r.h.s. of the last equation runs
over all neighbors of node $j$ excluding
vertex $i$. It is convenient to rewrite Eq.~(\ref{eq:site_sparse2})
as $\ln{(1 - r_{i \to j} / p)} =   \sum_k A_{jk} \, \ln{(1 -  r_{j \to k})}
- A_{ji}  \ln{(1 -  r_{j \to i})}$.
Defining the vectors $\vec{w}$ and
$\vec{t}$ such that their $(i \to j)$-th components
are respectively given by $w_{i \to j} = \ln(1 - r_{i \to j})$
and  $z_{i \to j} = \ln(1 -  r_{i \to j} /p )$, the system
of equations~(\ref{eq:site_sparse2}) 
becomes
equivalent to the vectorial equation
\begin{equation}
  \vec{t} = M \, \vec{w} \;.
  \label{eq:site_sparse2_vec}
\end{equation}
The generic element of the $2E \times 2E$
square matrix $M$ is given by
\begin{equation}
M_{i \to j, k \to \ell} = \delta_{j,k} (1 - \delta_{i,\ell}) \; ,
\label{eq:nonback}
\end{equation}
where $\delta_{x,y}$ is the
Kronecker delta function defined as
$\delta_{x,y} =1$ if $x=y$, and
$\delta_{x,y} =0$, otherwise.
Thus, the generic entry of the matrix $M$ is different
from zero only if the ending node of the
edge $i \to j$ corresponds to the starting
vertex of the edge $k \to \ell$, but
the starting and ending nodes $i$ and $\ell$
are different. This matrix is known
as the non-backtracking matrix of the 
graph~\cite{hashimoto1989zeta, krzakala2013spectral}.
A trivial solution of the preceding equation
is given by $\vec{r}=\vec{0}$, which in turn
leads to $\vec{s}=\vec{0}$. In proximity of this
configuration, we can still make use of
the truncated Taylor expansion of the logarithm,
and rewrite Eqs.~(\ref{eq:site_sparse1})
and ~(\ref{eq:site_sparse2_vec})
respectively as
\begin{equation}
  s_{i} = p \sum_j A_{ij} r_{i \to j} \quad  \textrm{and} \quad \vec{r} = p \, M \, \vec{r} \;.
  \label{eq:stability_site_sparse}
\end{equation}
Using arguments similar to those applied
to Eq.~(\ref{eq:stability_site_dense}),
we can say that, according to
Eq.~(\ref{eq:stability_site_sparse}), the percolation
threshold equals $p_c = 1/\mu$, with
$\mu$ principal eigenvalue of the
non-backtracking matrix of the graph, and that
slightly on the right of the critical point the probability
$s_i$ grows linearly with the sum of the components of the
principal eigenvector of the
non-backtracking matrix corresponding to edges pointing out
from node $i$. The entire percolation diagram
can be instead obtained by numerically
solving the system of Eqs.~(\ref{eq:site_sparse1})
and ~(\ref{eq:site_sparse2}).

To summarize, the results presented so far tell
us two main interesting things.
First, the difference between the
two approaches resides only in the
inclusion or exclusion of self-reinforcement
effects among local variables. In this sense,
Eqs.~(\ref{eq:site_sparse1}) and~(\ref{eq:site_sparse2})
represent an improvement to Eq.~(\ref{eq:site_dense}),
but both approaches are based on the same
principles and approximations. 
This first observation serves to reunite recent
predictions on percolation thresholds under the
same theory~\cite{bollobas2010percolation, PhysRevLett.113.208702,
PhysRevLett.113.208701}. 
Second, the way in which individual probabilities behave slightly
on the right of the critical point
allow us to understand why the prediction
of the percolation threshold of
Eq.~(\ref{eq:stability_site_sparse}) may become
inaccurate in networks with localized eigenstates
of the non-backtracking matrix~\cite{radicchi2014predicting}.

Next, we propose the generalization
of the previous equations to describe
percolation transitions in two interdependent
networks. 
Indicate with $A$ and $B$ the adjacency matrices of the
two network layers. Our first attempt to write the
the probability $s_i$ that node $i$ is in the 
largest mutually connected cluster of the system is given by
\begin{equation}
  s_i = p \; [ S_{\mathcal{AB}_i} + (1- S_{\mathcal{AB}_i}) \; S_{\mathcal{A-B}_i} \;  S_{\mathcal{B-A}_i} ]  \, ,
\label{eq:interd_dense}  
\end{equation}
where $S_{\mathcal{X}} = 1  - \prod_{j \in \mathcal{X}} (1 - s_j)$
is the probability that at least one of the nodes $j$
in the set $\mathcal{X}$ is part of the largest cluster
(for the empty set $\emptyset$, we
have $S_{\emptyset} = 0$). In the definition
of Eq.~(\ref{eq:interd_dense}), we have
implicitly defined three disjoint
sets of nodes: $\mathcal{AB}_i = \mathcal{N}^{A}_i \cap \mathcal{N}^{B}_i$ 
is the set of nodes that are neighbors of vertex $i$ in both
layers, $\mathcal{A-B}_i = \mathcal{N}^{A}_i \setminus \mathcal{AB}_i$
is the set of nodes connected to vertex $i$ only in layer $A$
but not in $B$, and
$\mathcal{B-A}_i = \mathcal{N}^{B}_i \setminus \mathcal{AB}_i$ is
the set of nodes that are neighbors of vertex $i$ in layer $B$
but not in $A$. 
Eq.~(\ref{eq:interd_dense}) essentially states
that, given that the vertex is occupied,
the probability $s_i$ for node $i$ of being part
of the largest mutually connected cluster is
given by the sum of two contributions:
(i) the probability to be connected to the
largest cluster thanks to at least
one vertex that is connected to $i$ in both layers;
(ii) if the latter condition is not true, the probability
that node $i$ is connected to the largest cluster through
at least one node $k$ in layer $A$ and one node $\ell$ in
layer $B$, with $k \neq \ell$.
Note that, if the network layers are identical, then
Eq.~(\ref{eq:interd_dense}) correctly reduces to
Eq.~(\ref{eq:site_dense}).
In other terms, one can
split the set of edges in the system in three
different subset, and then construct three
different graphs on the basis of this unique division (see Fig.~1):
the intersection graph with adjacency matrix
given by the Hadamard product of the matrices $A$ and
$B$ [i.e., the $(i,j)$-th element of the
  adjacency matrix is $A_{ij}B_{ij}$]; the remnant of
network $A$, where edges between nodes $i$ and $j$
are present only if $A_{ij}(1-B_{ij})=1$; the remainder of graph $B$,
whose $(i,j)$-th adjacency matrix element equals $B_{ij}(1-A_{ij})$.
If we make use of the
vector $\vec{u}$ previously defined, we can
write $S_{\mathcal{AB}_i} = 1 - \exp{[\sum_j A_{ij} B_{ij} u_j]}$,
$S_{\mathcal{A - B}_i} = 1 - \exp{[\sum_j A_{ij} (1-B_{ij}) u_j]}$
and $S_{\mathcal{B - A}_i} = 1 - \exp{[\sum_j B_{ij} (1-A_{ij}) u_j]}$.
Thus, the numerical solution of
Eq.~(\ref{eq:interd_dense}) can be obtained 
in a certain number of iterations, each having
a computational complexity that grows at maximum as the
number of edges present in the denser layer. Unfortunately,
the Taylor expansion of the r.h.s. of
Eq.~(\ref{eq:interd_dense})
gives us only some insights
about the structure of the solution, but it does not allow
to reduce the original problem to a simple eigenvalue/eigenvector
equation as in the case of isolated networks
(see Appendix).

\begin{figure*}[!htb]
\begin{center}  
\includegraphics[width=0.75\textwidth]{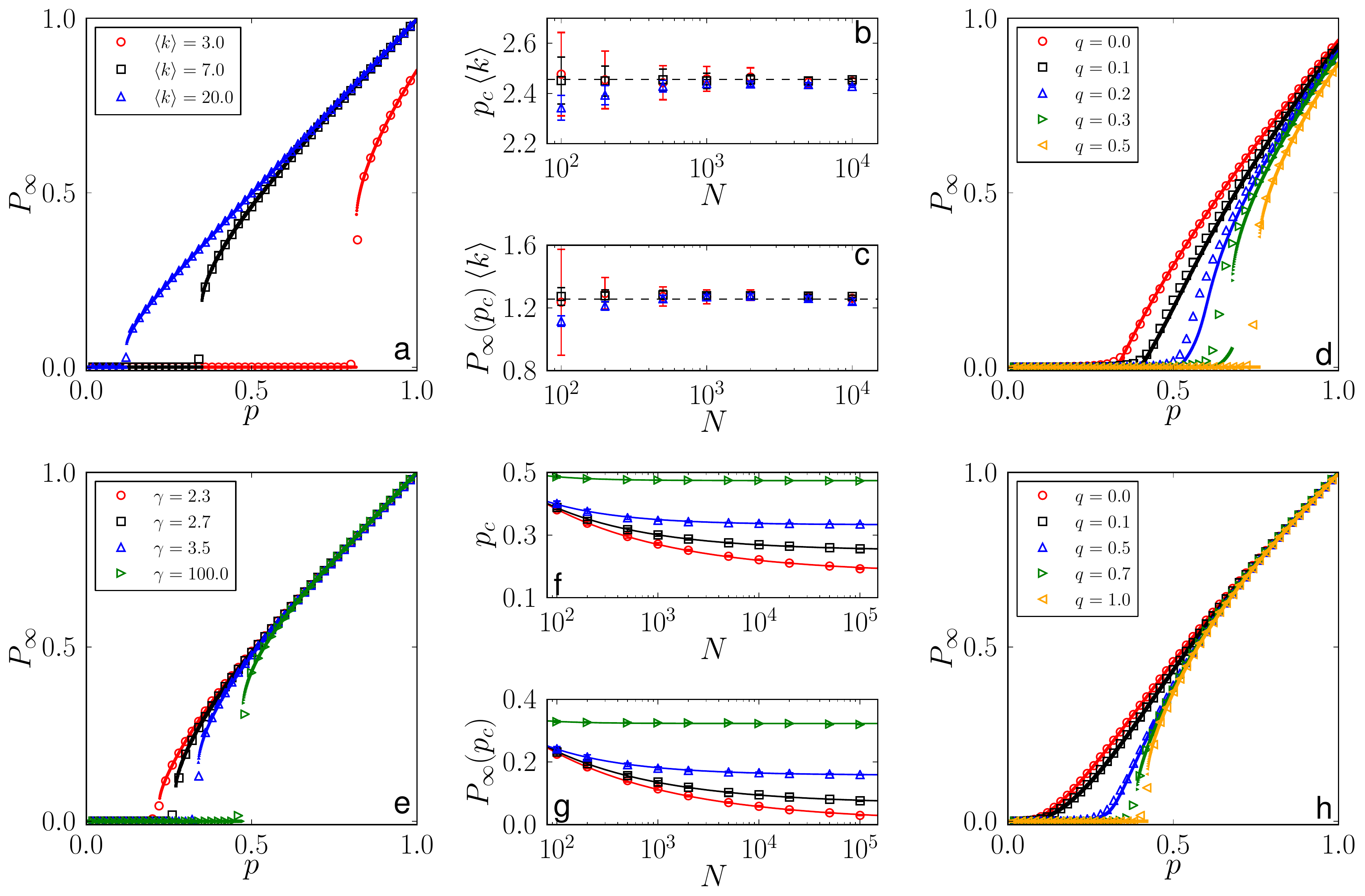}                                                                                 
\end{center}                                                                                                                     
  \caption{
Percolation transition in artificial interconnected
    networks. {\bf a)} Percolation diagrams for interdependent
    Erd\H{o}s-R\'enyi graphs. We generate network layers
    with the same average
    degree $\langle k \rangle$, and compare results of
    numerical simulations (large symbols) with the
    solutions of our equations (small symbols). 
    Results are obtained on a single
    instance of the network model, where both layers have
    size $N=10^4$. Different colors and symbols
    refer to different values of the average degree $\langle k \rangle$.
    {\bf b)} and {\bf c)} For a given size
    of the network layers, we generate
    several instances of the model, and
    compute the percolation threshold $p_c$ and the height
    of the order parameter at criticality $P_\infty(p_c)$.
    Both quantities are multiplied by the average degree $\langle k \rangle$.
    Points refer to average values obtained over several realizations
    of the graph model, while
    error bars stand for standard deviations. We use the same
    symbols and colors as those of panel a.
    Results are compared with
    the theoretical expected values $p_c \langle k \rangle = 2.4554$
    and $P_\infty(p_c) \langle k \rangle = 1.2564$ (dashed black lines).
    {\bf d)} We generate a single Erd\H{o}s-R\'enyi graph
    with $N = 10^4$ and $\langle k \rangle = 3.0$, and use it
    as structure for both layers. We then exchange with probability
    $q$ the label of 
    every node of layer $B$ with a randomly selected vertex.
    {\bf e)} Percolation diagrams for single instances
    of interconnected scale-free networks. Each layer is obtained
    by randomly connecting vertices whose degrees obey the
    distribution $P(k) \sim k^{-\gamma}$, if $k \in [5, \sqrt{N}]$, and
    $P(k)  =0$, otherwise. Here, $N = 10^4$.
    {\bf f)} and {\bf g)} We compute  the average values of $p_c$
    and  $P_\infty(p_c)$
    in several realizations of model composed of 
    interdependent scale-free networks with size $N$. 
    Standard deviations have size comparable with those of the symbol
    sizes. We use the same symbol/color scheme as in panel e.
    Full lines in panel f stand for best estimates of fits
    of empirical points with the function $p_c(N) = p_c + N^{-\alpha}$.
    The same type of functions are used in panel g to extrapolate the 
    asymptotic value of
    $P_\infty(p_c)$. For any value of the
    degree exponent, we find that the asymptotic values 
    of both quantities 
    are strictly larger than zero (see Appendix). {\bf h)} We generate a graph
    with $N = 10^4$ nodes, and degrees extracted 
    from a power-law distribution with exponent $\gamma = 2.5$ 
    and support $[3, \sqrt{N}]$.
    We use this network as structure for both layers. 
    We then exchange with probability
    $q$ the label of 
    every node of layer $B$ with a randomly selected vertex.
  }
\label{fig:2}
\end{figure*}

Also here, we can avoid
the presence of self-reinforcing mechanisms
among variables by excluding already visited edges.
The equations read as
\begin{equation}
  s_i = p \; [ R_{\mathcal{AB}_i} + (1- R_{\mathcal{AB}_i}) \; R_{\mathcal{A-B}_i} \;  R_{\mathcal{B-A}_i} ]  \, ,
\label{eq:interd_sparse1}  
\end{equation}
and 
\begin{equation}
  r_{i \to j} = p \; [ R_{\mathcal{AB}_{j} \setminus \{i\}} + (1- R_{\mathcal{AB}_{j} \setminus \{i\}}) \; R_{\mathcal{A-B}_{j} \setminus \{i\}} \;  R_{\mathcal{B-A}_{j} \setminus \{i\}} ]  \, ,
\label{eq:interd_sparse2}  
\end{equation}
with $R_{\mathcal{X}_i} = 1 - \prod_{j \in \mathcal{X}} \, (1 - r_{i \to j})$ , and
the three sets $\mathcal{AB}_i$, $\mathcal{A-B}_i$
and $\mathcal{B-A}_i$ are
defined as above. If the network layers are identical, then
Eqs.~(\ref{eq:interd_sparse1}) and~(\ref{eq:interd_sparse2})
reduce to
Eqs.~(\ref{eq:site_sparse1}) and~(\ref{eq:site_sparse2}).
If we indicate with $\vec{r}^{(AB)}$ 
the vector whose components
are generated by edges present in the
intersection graph, $\vec{w}^{(AB)}$ the vector
with entries of the type
$\vec{w}^{(AB)}_{i \to j} = \ln(1 - \vec{r}^{(AB)}_{i \to j} )$,
and $M^{(AB)}$ the non-backtracking matrix
obtained from the adjacency matrix
of intersection between layers, we can write
$R_{\mathcal{AB}_{j} \setminus \{i\}} = 1 - \exp{[M^{(AB)} \, \vec{w}^{(AB)}]}$.
In a similar spirit, we can also write
$R_{\mathcal{A-B}_{j} \setminus \{i\}} = 1 - \exp{[M^{(A-B)} \, \vec{w}^{(A-B)}]}$
and $R_{\mathcal{B-A}_{j} \setminus \{i\}} = 1 - \exp{[M^{(B-A)} \, \vec{w}^{(B-A)}]}$,
where these equations are valid only for edges that belong
to either layer $A$ or layer $B$.
Obtaining a numerical
solution of Eqs.~(\ref{eq:interd_sparse1})
and~(\ref{eq:interd_sparse2}) by iteration is thus
relatively fast, since every iteration
has a computational complexity at maximum equal
to twice the number of edges present in 
the denser network layer. This is
a great achievement given the high complexity
of the algorithm necessary to draw the
phase diagram for the percolation
process in interdependent networks by means of
direct numerical simulations~\cite{hwang2014efficient}.

\begin{figure}
\begin{center}  
\includegraphics[width=0.45\textwidth]{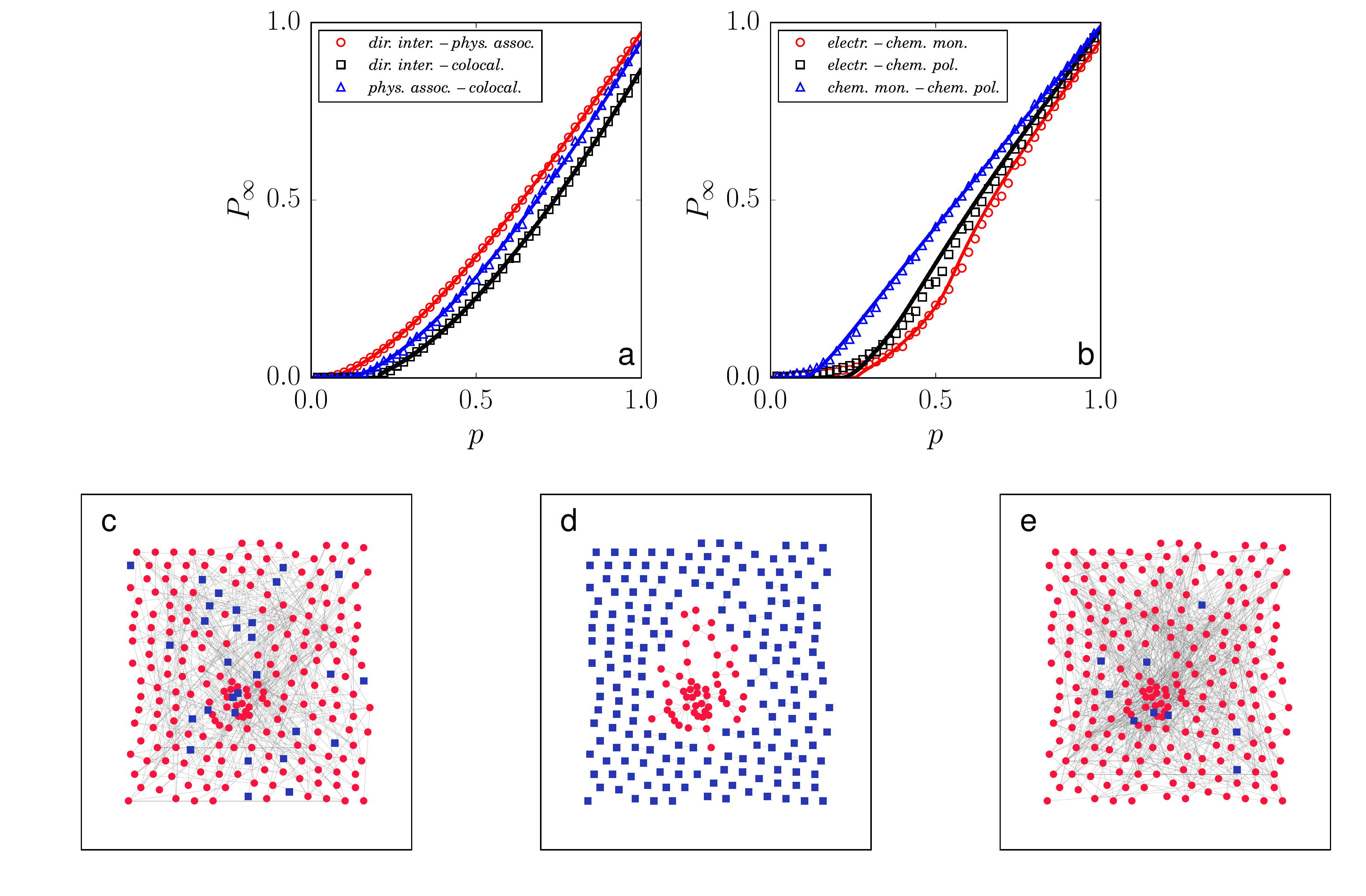}                                                                                 
\end{center}                                                                                                                     
  \caption{
Percolation transition in interdependent biological
    networks. {\bf a)} Phase diagram for the multilayer {\it H. sapiens}
    protein interaction network~\cite{stark2006biogrid, de2014muxviz}.
    Edges in different layers represent diverse type of 
    connections among proteins:
    direct interaction, physical association, and colocalization.
    When analyzing a multiplex with two of these layers, we restrict
    our attention only on the set of nodes present in both layers.
    For each of the three systems formed by two
    interconnected networks that we can generate
    with this data, we draw the percolation diagram
    by means of numerical simulations (large symbols)
    and numerical solution of our equations (small symbols).
    {\bf b)} Phase diagram for the multilayer network 
    of the {\it C. elegans}
    connectome~\cite{de2014muxviz}. Edges in different
    layers represent different types of synaptic 
    junctions among the neurons: electrical, 
    chemical monadic, and chemical polyadic.
    {\bf c)} Decomposition of the multilayer {\it C. elegans}
    connectome. Remnant of the layer corresponding to
    electrical junctions, {\bf d)} intersection among the
    layers corresponding to electrical and chemical monadic
    interactions and {\bf e)} remainder of the layer corresponding to
    chemical monadic junctions. In the various panels, 
    nodes belonging to the largest connected component
    are visualized with red circles. All other nodes are instead
    represented with blue squares.
  }
\label{fig:3}
\end{figure}

\begin{figure}
\begin{center}  
\includegraphics[width=0.45\textwidth]{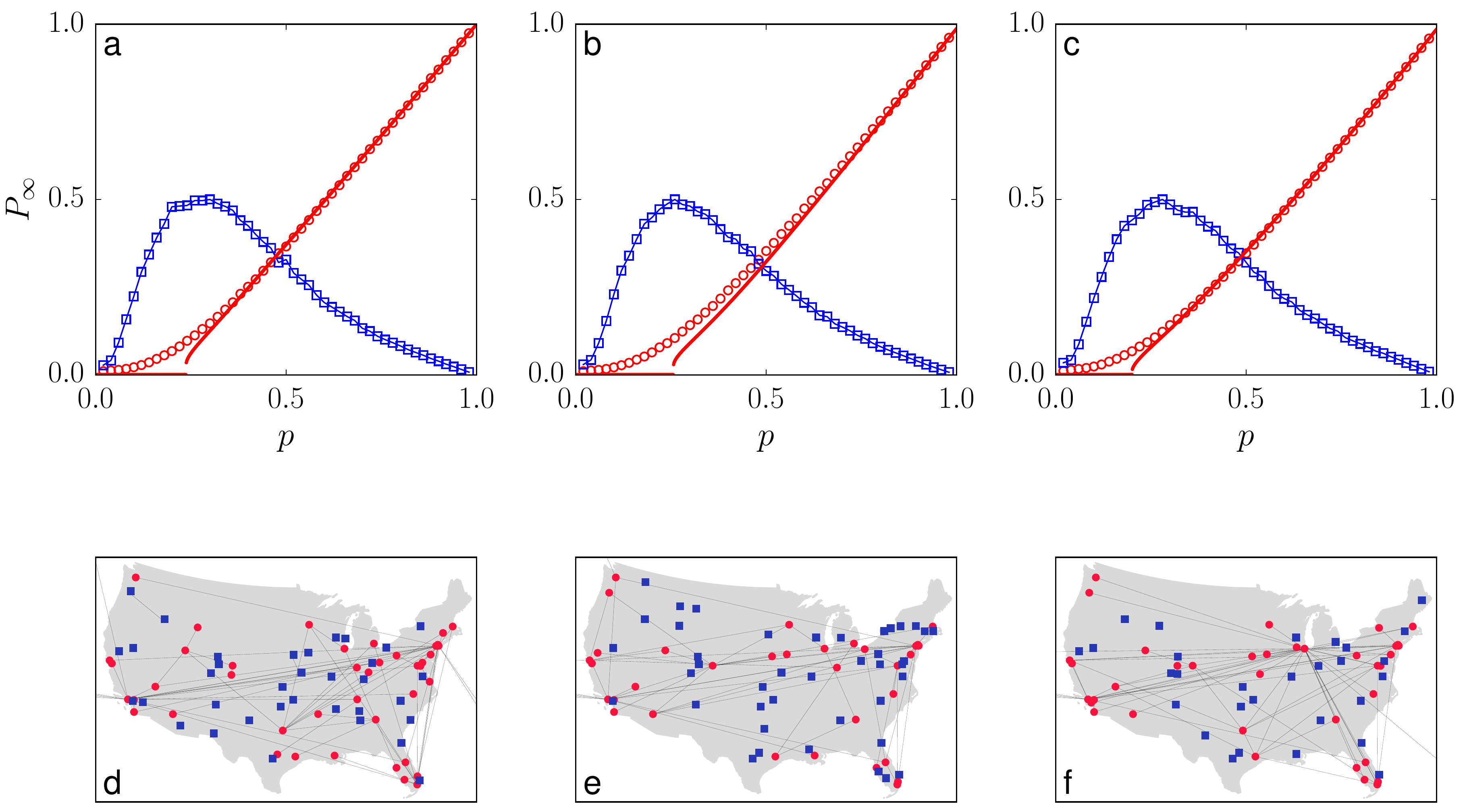}                                                                                 
\end{center}                                                                                                                     
\caption{
Percolation transition in interconnected transportation
networks. {\bf a)} The system is obtained by combining
Delta and American Airlines routes.  
We consider only US domestic flights 
operated in January, 2014~\cite{flights}, and
construct an interconnected network 
where airports are nodes, 
and connections on the layers
are determined by the existence of at least
a flight between the two locations.
In the percolation diagram, 
large red circles are 
results of numerical simulations, whereas small red circles
represent the solutions of our equations. Blue squares
represent
susceptibility, a measure of the
fluctuation across realizations of the
percolation model, whose peak location is often used as
a proxy for the identification of the 
critical threshold $p_c$.
{\bf b)} Same as in a, but for the combination 
of Delta and United flights. {\bf c)} Same as in a, 
but for the combination of American Airlines and United flights.
{\bf d, e, and f)} Intersection graphs for the systems analyzed
respectively in panels a, b and c.  In the various network
visualizations,  nodes belonging to the 
largest connected component
are visualized with red circles. All other nodes are instead
represented with blue squares.
  }
\label{fig:4}
\end{figure}

Phase diagrams obtained through the numerical
solution of Eqs.~(\ref{eq:interd_sparse1})
and~(\ref{eq:interd_sparse2}) reproduce the
results of numerical simulations
very accurately. In Fig.~2a for example,
we consider systems composed
of two independent Erd\H{o}s-R\'enyi network
models with different values of the average
degree $\langle k \rangle$, where each 
network layer is generated
by connecting pairs of vertices with probability
$\langle k \rangle / N$.
A fundamental feature that the diagrams 
reveal is the presence of a sudden jump in the
order parameter $P_\infty$ at
a certain threshold $p_c$. We stress that our equations
predict the existence of
first-order percolation transitions
in networks of finite size, and not just
in the thermodynamic limit. As Figs.~2b and c show,
the location of the critical point $p_c$, and the
height of the jump of the order parameter
are well described by predictions valid
for this type of graph models
in the limit of infinite size\cite{buldyrev2010catastrophic,
  gao2012networks, son2012percolation}.
We argue that a sudden jump
in the order parameter is present only
if the contribution of the remainders 
dominates the importance of the
intersection. This condition is certainly
verified in interdependent Erd\H{o}s-R\'enyi graphs, 
where the intersection
is composed of a very small number of edges,
roughly equal to  $\langle k \rangle / 2$, while
the number of edges in each remnants is 
proportional to $\langle k \rangle \, N /2$.
Our intuition is fully supported by the
results of Fig.~2d. Here, we control for the
weight of the intersection with respect to those of
the remnants in a simple fashion~\cite{bianconi2014percolation}.
The two layers
are given by exactly the same network structure.
Indices of interdependent nodes are however shuffled
with a given probability $q$. As $q$ grows, the
percolation transition changes its features: we pass from a 
continuous phase 
transition for small values of $q$, through a mix between
a second- and a first-order structural change at intermediate
values of $q$, to a discontinuous phase transition
for sufficiently large values of $q$.
The same type of considerations hold
when network layers are scale-free random graphs.
The transition is always discontinuous if the
layers are uncorrelated, so that
only remainders are present (Fig.~2e). This
can be viewed by the existence of a finite
value of the critical threshold $p_c$ (Fig.~2f),
and a jump of non null height of the
order parameter at criticality (Fig.~2g). Still, the
nature of the phase transition can be
tuned from first to second order by
simply decreasing the density of the intersection
relatively to those of the remainders (Fig.~2h).
Our argument about the 
dependence of the nature of the transition
on the weight of the intersection compared to
those of the remnants may serve to explain
why real interdependent networks are
not exposed to catastrophic failures~\cite{reis2014avoiding}.
In Fig.~3, we draw the percolation
diagrams for two interconnected systems
of interest in the biological sciences:
the {\it H. sapiens}
protein interaction network~\cite{stark2006biogrid, de2014muxviz}, and
the {\it C. elegans} connectome~\cite{de2014muxviz}.
Both these interconnected systems undergo
continuous percolation transitions.
Interestingly, this behavior is not
caused by the amount of redundancy
among layers, but rather the ``quality''
of the edges shared across layers. Connections
in the intersection graph account, in fact, for less than
$10\%$ in five out of the six interdependent
networks analyzed here. It seems therefore that 
these organisms
have developed interconnected networks
sharing a core of ``high quality'' edges 
to prevent catastrophic failures.
Whereas the robustness we observe in biological networks
can be viewed as the result of a selective
evolutionary process, one may argue that
man-made interdependent systems could have been
instead not perfectly designed to resist
to random damages of their components.
This is indeed what arise from
the analysis of the multilayer
air transportation network within the
US (Fig.~4)~\cite{guimera2005worldwide, colizza2006role}.
The system shows fragility, with a
sudden jump of the order parameter.
On the other hand, the height of the
jump is not as dramatic as observed in random
uncorrelated graphs. Major airports all belong
to the largest connected component of
the intersection graph, and their
connections constitute a set of
high quality edges that avoid truly catastrophic
changes in the connectedness of the entire
interdependent system. From these examples, 
it seems therefore that real
interdependent networks may be not
so fragile as previously believed.


\begin{thebibliography}{30}
\expandafter\ifx\csname natexlab\endcsname\relax\def\natexlab#1{#1}\fi
\expandafter\ifx\csname bibnamefont\endcsname\relax
  \def\bibnamefont#1{#1}\fi
\expandafter\ifx\csname bibfnamefont\endcsname\relax
  \def\bibfnamefont#1{#1}\fi
\expandafter\ifx\csname citenamefont\endcsname\relax
  \def\citenamefont#1{#1}\fi
\expandafter\ifx\csname url\endcsname\relax
  \def\url#1{\texttt{#1}}\fi
\expandafter\ifx\csname urlprefix\endcsname\relax\def\urlprefix{URL }\fi
\providecommand{\bibinfo}[2]{#2}
\providecommand{\eprint}[2][]{\url{#2}}

\bibitem[{\citenamefont{Stauffer and Aharony}(1991)}]{stauffer1991introduction}
\bibinfo{author}{\bibfnamefont{D.}~\bibnamefont{Stauffer}} \bibnamefont{and}
  \bibinfo{author}{\bibfnamefont{A.}~\bibnamefont{Aharony}},
  \emph{\bibinfo{title}{Introduction to percolation theory}}
  (\bibinfo{publisher}{Taylor and Francis}, \bibinfo{year}{1991}).

\bibitem[{\citenamefont{Kirkpatrick}(1973)}]{kirkpatrick1973percolation}
\bibinfo{author}{\bibfnamefont{S.}~\bibnamefont{Kirkpatrick}},
  \bibinfo{journal}{Reviews of modern physics} \textbf{\bibinfo{volume}{45}},
  \bibinfo{pages}{574} (\bibinfo{year}{1973}).

\bibitem[{\citenamefont{Berkowitz}(1995)}]{berkowitz1995analysis}
\bibinfo{author}{\bibfnamefont{B.}~\bibnamefont{Berkowitz}},
  \bibinfo{journal}{Mathematical Geology} \textbf{\bibinfo{volume}{27}},
  \bibinfo{pages}{467} (\bibinfo{year}{1995}).

\bibitem[{\citenamefont{Pastor-Satorras and
  Vespignani}(2001)}]{pastor2001epidemic}
\bibinfo{author}{\bibfnamefont{R.}~\bibnamefont{Pastor-Satorras}}
  \bibnamefont{and}
  \bibinfo{author}{\bibfnamefont{A.}~\bibnamefont{Vespignani}},
  \bibinfo{journal}{Physical review letters} \textbf{\bibinfo{volume}{86}},
  \bibinfo{pages}{3200} (\bibinfo{year}{2001}).

\bibitem[{\citenamefont{Newman}(2002)}]{newman2002spread}
\bibinfo{author}{\bibfnamefont{M.~E.} \bibnamefont{Newman}},
  \bibinfo{journal}{Physical review E} \textbf{\bibinfo{volume}{66}},
  \bibinfo{pages}{016128} (\bibinfo{year}{2002}).

\bibitem[{\citenamefont{Albert et~al.}(2000)\citenamefont{Albert, Jeong, and
  Barab{\'a}si}}]{albert2000error}
\bibinfo{author}{\bibfnamefont{R.}~\bibnamefont{Albert}},
  \bibinfo{author}{\bibfnamefont{H.}~\bibnamefont{Jeong}}, \bibnamefont{and}
  \bibinfo{author}{\bibfnamefont{A.-L.} \bibnamefont{Barab{\'a}si}},
  \bibinfo{journal}{Nature} \textbf{\bibinfo{volume}{406}},
  \bibinfo{pages}{378} (\bibinfo{year}{2000}).

\bibitem[{\citenamefont{Cohen et~al.}(2000)\citenamefont{Cohen, Erez,
  Ben-Avraham, and Havlin}}]{cohen2000resilience}
\bibinfo{author}{\bibfnamefont{R.}~\bibnamefont{Cohen}},
  \bibinfo{author}{\bibfnamefont{K.}~\bibnamefont{Erez}},
  \bibinfo{author}{\bibfnamefont{D.}~\bibnamefont{Ben-Avraham}},
  \bibnamefont{and} \bibinfo{author}{\bibfnamefont{S.}~\bibnamefont{Havlin}},
  \bibinfo{journal}{Physical review letters} \textbf{\bibinfo{volume}{85}},
  \bibinfo{pages}{4626} (\bibinfo{year}{2000}).

\bibitem[{\citenamefont{Callaway et~al.}(2000)\citenamefont{Callaway, Newman,
  Strogatz, and Watts}}]{callaway2000network}
\bibinfo{author}{\bibfnamefont{D.~S.} \bibnamefont{Callaway}},
  \bibinfo{author}{\bibfnamefont{M.~E.} \bibnamefont{Newman}},
  \bibinfo{author}{\bibfnamefont{S.~H.} \bibnamefont{Strogatz}},
  \bibnamefont{and} \bibinfo{author}{\bibfnamefont{D.~J.} \bibnamefont{Watts}},
  \bibinfo{journal}{Physical review letters} \textbf{\bibinfo{volume}{85}},
  \bibinfo{pages}{5468} (\bibinfo{year}{2000}).

\bibitem[{\citenamefont{Dorogovtsev et~al.}(2008)\citenamefont{Dorogovtsev,
  Goltsev, and Mendes}}]{dorogovtsev2008critical}
\bibinfo{author}{\bibfnamefont{S.~N.} \bibnamefont{Dorogovtsev}},
  \bibinfo{author}{\bibfnamefont{A.~V.} \bibnamefont{Goltsev}},
  \bibnamefont{and} \bibinfo{author}{\bibfnamefont{J.~F.}
  \bibnamefont{Mendes}}, \bibinfo{journal}{Reviews of Modern Physics}
  \textbf{\bibinfo{volume}{80}}, \bibinfo{pages}{1275} (\bibinfo{year}{2008}).

\bibitem[{\citenamefont{Buldyrev et~al.}(2010)\citenamefont{Buldyrev, Parshani,
  Paul, Stanley, and Havlin}}]{buldyrev2010catastrophic}
\bibinfo{author}{\bibfnamefont{S.~V.} \bibnamefont{Buldyrev}},
  \bibinfo{author}{\bibfnamefont{R.}~\bibnamefont{Parshani}},
  \bibinfo{author}{\bibfnamefont{G.}~\bibnamefont{Paul}},
  \bibinfo{author}{\bibfnamefont{H.~E.} \bibnamefont{Stanley}},
  \bibnamefont{and} \bibinfo{author}{\bibfnamefont{S.}~\bibnamefont{Havlin}},
  \bibinfo{journal}{Nature} \textbf{\bibinfo{volume}{464}},
  \bibinfo{pages}{1025} (\bibinfo{year}{2010}).

\bibitem[{\citenamefont{Szell et~al.}(2010)\citenamefont{Szell, Lambiotte, and
  Thurner}}]{szell2010multirelational}
\bibinfo{author}{\bibfnamefont{M.}~\bibnamefont{Szell}},
  \bibinfo{author}{\bibfnamefont{R.}~\bibnamefont{Lambiotte}},
  \bibnamefont{and} \bibinfo{author}{\bibfnamefont{S.}~\bibnamefont{Thurner}},
  \bibinfo{journal}{Proceedings of the National Academy of Sciences USA}
  \textbf{\bibinfo{volume}{107}}, \bibinfo{pages}{13636}
  (\bibinfo{year}{2010}).

\bibitem[{\citenamefont{Barth{\'e}lemy}(2011)}]{barthelemy2011spatial}
\bibinfo{author}{\bibfnamefont{M.}~\bibnamefont{Barth{\'e}lemy}},
  \bibinfo{journal}{Physics Reports} \textbf{\bibinfo{volume}{499}},
  \bibinfo{pages}{1} (\bibinfo{year}{2011}).

\bibitem[{\citenamefont{De~Domenico
  et~al.}(2014{\natexlab{a}})\citenamefont{De~Domenico, Sol{\'e}-Ribalta,
  G{\'o}mez, and Arenas}}]{de2014navigability}
\bibinfo{author}{\bibfnamefont{M.}~\bibnamefont{De~Domenico}},
  \bibinfo{author}{\bibfnamefont{A.}~\bibnamefont{Sol{\'e}-Ribalta}},
  \bibinfo{author}{\bibfnamefont{S.}~\bibnamefont{G{\'o}mez}},
  \bibnamefont{and} \bibinfo{author}{\bibfnamefont{A.}~\bibnamefont{Arenas}},
  \bibinfo{journal}{Proceedings of the National Academy of Sciences USA}
  \textbf{\bibinfo{volume}{111}}, \bibinfo{pages}{8351}
  (\bibinfo{year}{2014}{\natexlab{a}}).

\bibitem[{\citenamefont{Gao et~al.}(2012)\citenamefont{Gao, Buldyrev, Stanley,
  and Havlin}}]{gao2012networks}
\bibinfo{author}{\bibfnamefont{J.}~\bibnamefont{Gao}},
  \bibinfo{author}{\bibfnamefont{S.~V.} \bibnamefont{Buldyrev}},
  \bibinfo{author}{\bibfnamefont{H.~E.} \bibnamefont{Stanley}},
  \bibnamefont{and} \bibinfo{author}{\bibfnamefont{S.}~\bibnamefont{Havlin}},
  \bibinfo{journal}{Nature physics} \textbf{\bibinfo{volume}{8}},
  \bibinfo{pages}{40} (\bibinfo{year}{2012}).

\bibitem[{\citenamefont{Son et~al.}(2012)\citenamefont{Son, Bizhani,
  Christensen, Grassberger, and Paczuski}}]{son2012percolation}
\bibinfo{author}{\bibfnamefont{S.-W.} \bibnamefont{Son}},
  \bibinfo{author}{\bibfnamefont{G.}~\bibnamefont{Bizhani}},
  \bibinfo{author}{\bibfnamefont{C.}~\bibnamefont{Christensen}},
  \bibinfo{author}{\bibfnamefont{P.}~\bibnamefont{Grassberger}},
  \bibnamefont{and} \bibinfo{author}{\bibfnamefont{M.}~\bibnamefont{Paczuski}},
  \bibinfo{journal}{EPL (Europhysics Letters)} \textbf{\bibinfo{volume}{97}},
  \bibinfo{pages}{16006} (\bibinfo{year}{2012}).

\bibitem[{\citenamefont{Radicchi}(2014)}]{PhysRevX.4.021014}
\bibinfo{author}{\bibfnamefont{F.}~\bibnamefont{Radicchi}},
  \bibinfo{journal}{Physical Review X} \textbf{\bibinfo{volume}{4}},
  \bibinfo{pages}{021014} (\bibinfo{year}{2014}).

\bibitem[{\citenamefont{Reis et~al.}(2014)\citenamefont{Reis, Hu, Babino,
  Andrade~Jr, Canals, Sigman, and Makse}}]{reis2014avoiding}
\bibinfo{author}{\bibfnamefont{S.~D.} \bibnamefont{Reis}},
  \bibinfo{author}{\bibfnamefont{Y.}~\bibnamefont{Hu}},
  \bibinfo{author}{\bibfnamefont{A.}~\bibnamefont{Babino}},
  \bibinfo{author}{\bibfnamefont{J.~S.} \bibnamefont{Andrade~Jr}},
  \bibinfo{author}{\bibfnamefont{S.}~\bibnamefont{Canals}},
  \bibinfo{author}{\bibfnamefont{M.}~\bibnamefont{Sigman}}, \bibnamefont{and}
  \bibinfo{author}{\bibfnamefont{H.~A.} \bibnamefont{Makse}},
  \bibinfo{journal}{Nature Physics} \textbf{\bibinfo{volume}{10}},
  \bibinfo{pages}{762} (\bibinfo{year}{2014}).

\bibitem[{\citenamefont{Bollob{\'a}s et~al.}(2010)\citenamefont{Bollob{\'a}s,
  Borgs, Chayes, Riordan et~al.}}]{bollobas2010percolation}
\bibinfo{author}{\bibfnamefont{B.}~\bibnamefont{Bollob{\'a}s}},
  \bibinfo{author}{\bibfnamefont{C.}~\bibnamefont{Borgs}},
  \bibinfo{author}{\bibfnamefont{J.}~\bibnamefont{Chayes}},
  \bibinfo{author}{\bibfnamefont{O.}~\bibnamefont{Riordan}},
  \bibnamefont{et~al.}, \bibinfo{journal}{The Annals of Probability}
  \textbf{\bibinfo{volume}{38}}, \bibinfo{pages}{150} (\bibinfo{year}{2010}).

\bibitem[{\citenamefont{Karrer et~al.}(2014)\citenamefont{Karrer, Newman, and
  Zdeborov\'a}}]{PhysRevLett.113.208702}
\bibinfo{author}{\bibfnamefont{B.}~\bibnamefont{Karrer}},
  \bibinfo{author}{\bibfnamefont{M.~E.~J.} \bibnamefont{Newman}},
  \bibnamefont{and}
  \bibinfo{author}{\bibfnamefont{L.}~\bibnamefont{Zdeborov\'a}},
  \bibinfo{journal}{Physical Review Letters} \textbf{\bibinfo{volume}{113}},
  \bibinfo{pages}{208702} (\bibinfo{year}{2014}).

\bibitem[{\citenamefont{Hamilton and Pryadko}(2014)}]{PhysRevLett.113.208701}
\bibinfo{author}{\bibfnamefont{K.~E.} \bibnamefont{Hamilton}} \bibnamefont{and}
  \bibinfo{author}{\bibfnamefont{L.~P.} \bibnamefont{Pryadko}},
  \bibinfo{journal}{Physical Review Letters} \textbf{\bibinfo{volume}{113}},
  \bibinfo{pages}{208701} (\bibinfo{year}{2014}).

\bibitem[{\citenamefont{Hashimoto}(1989)}]{hashimoto1989zeta}
\bibinfo{author}{\bibfnamefont{K.-i.} \bibnamefont{Hashimoto}},
  \bibinfo{journal}{Automorphic forms and geometry of arithmetic varieties.}
  pp. \bibinfo{pages}{211--280} (\bibinfo{year}{1989}).

\bibitem[{\citenamefont{Krzakala et~al.}(2013)\citenamefont{Krzakala, Moore,
  Mossel, Neeman, Sly, Zdeborov{\'a}, and Zhang}}]{krzakala2013spectral}
\bibinfo{author}{\bibfnamefont{F.}~\bibnamefont{Krzakala}},
  \bibinfo{author}{\bibfnamefont{C.}~\bibnamefont{Moore}},
  \bibinfo{author}{\bibfnamefont{E.}~\bibnamefont{Mossel}},
  \bibinfo{author}{\bibfnamefont{J.}~\bibnamefont{Neeman}},
  \bibinfo{author}{\bibfnamefont{A.}~\bibnamefont{Sly}},
  \bibinfo{author}{\bibfnamefont{L.}~\bibnamefont{Zdeborov{\'a}}},
  \bibnamefont{and} \bibinfo{author}{\bibfnamefont{P.}~\bibnamefont{Zhang}},
  \bibinfo{journal}{Proceedings of the National Academy of Sciences}
  \textbf{\bibinfo{volume}{110}}, \bibinfo{pages}{20935}
  (\bibinfo{year}{2013}).

\bibitem[{\citenamefont{Radicchi}(2015)}]{radicchi2014predicting}
\bibinfo{author}{\bibfnamefont{F.}~\bibnamefont{Radicchi}},
  \bibinfo{journal}{Physical Review E} \textbf{\bibinfo{volume}{91}},
  \bibinfo{pages}{010801} (\bibinfo{year}{2015}).

\bibitem[{\citenamefont{Hwang et~al.}(2014)\citenamefont{Hwang, Choi, Lee, and
  Kahng}}]{hwang2014efficient}
\bibinfo{author}{\bibfnamefont{S.}~\bibnamefont{Hwang}},
  \bibinfo{author}{\bibfnamefont{S.}~\bibnamefont{Choi}},
  \bibinfo{author}{\bibfnamefont{D.}~\bibnamefont{Lee}}, \bibnamefont{and}
  \bibinfo{author}{\bibfnamefont{B.}~\bibnamefont{Kahng}},
  \bibinfo{journal}{arXiv preprint arXiv:1409.1147}  (\bibinfo{year}{2014}).

\bibitem[{\citenamefont{Stark et~al.}(2006)\citenamefont{Stark, Breitkreutz,
  Reguly, Boucher, Breitkreutz, and Tyers}}]{stark2006biogrid}
\bibinfo{author}{\bibfnamefont{C.}~\bibnamefont{Stark}},
  \bibinfo{author}{\bibfnamefont{B.-J.} \bibnamefont{Breitkreutz}},
  \bibinfo{author}{\bibfnamefont{T.}~\bibnamefont{Reguly}},
  \bibinfo{author}{\bibfnamefont{L.}~\bibnamefont{Boucher}},
  \bibinfo{author}{\bibfnamefont{A.}~\bibnamefont{Breitkreutz}},
  \bibnamefont{and} \bibinfo{author}{\bibfnamefont{M.}~\bibnamefont{Tyers}},
  \bibinfo{journal}{Nucleic acids research} \textbf{\bibinfo{volume}{34}},
  \bibinfo{pages}{D535} (\bibinfo{year}{2006}).

\bibitem[{\citenamefont{De~Domenico
  et~al.}(2014{\natexlab{b}})\citenamefont{De~Domenico, Porter, and
  Arenas}}]{de2014muxviz}
\bibinfo{author}{\bibfnamefont{M.}~\bibnamefont{De~Domenico}},
  \bibinfo{author}{\bibfnamefont{M.~A.} \bibnamefont{Porter}},
  \bibnamefont{and} \bibinfo{author}{\bibfnamefont{A.}~\bibnamefont{Arenas}},
  \bibinfo{journal}{Journal of Complex Networks} p. \bibinfo{pages}{cnu038}
  (\bibinfo{year}{2014}{\natexlab{b}}).

\bibitem[{fli()}]{flights}
\emph{\bibinfo{title}{Bureau of transportation statistics}},
  \bibinfo{howpublished}{\url{http://www.transtats.bts.gov}},
  \bibinfo{note}{accessed: 2015-01-18}.

\bibitem[{\citenamefont{Bianconi and
  Dorogovtsev}(2014)}]{bianconi2014percolation}
\bibinfo{author}{\bibfnamefont{G.}~\bibnamefont{Bianconi}} \bibnamefont{and}
  \bibinfo{author}{\bibfnamefont{S.~N.} \bibnamefont{Dorogovtsev}},
  \bibinfo{journal}{arXiv preprint arXiv:1411.4160}  (\bibinfo{year}{2014}).

\bibitem[{\citenamefont{Guimer{\`a} et~al.}(2005)\citenamefont{Guimer{\`a},
  Mossa, Turtschi, and Amaral}}]{guimera2005worldwide}
\bibinfo{author}{\bibfnamefont{R.}~\bibnamefont{Guimer{\`a}}},
  \bibinfo{author}{\bibfnamefont{S.}~\bibnamefont{Mossa}},
  \bibinfo{author}{\bibfnamefont{A.}~\bibnamefont{Turtschi}}, \bibnamefont{and}
  \bibinfo{author}{\bibfnamefont{L.~N.} \bibnamefont{Amaral}},
  \bibinfo{journal}{Proceedings of the National Academy of Sciences USA}
  \textbf{\bibinfo{volume}{102}}, \bibinfo{pages}{7794} (\bibinfo{year}{2005}).

\bibitem[{\citenamefont{Colizza et~al.}(2006)\citenamefont{Colizza, Barrat,
  Barth{\'e}lemy, and Vespignani}}]{colizza2006role}
\bibinfo{author}{\bibfnamefont{V.}~\bibnamefont{Colizza}},
  \bibinfo{author}{\bibfnamefont{A.}~\bibnamefont{Barrat}},
  \bibinfo{author}{\bibfnamefont{M.}~\bibnamefont{Barth{\'e}lemy}},
  \bibnamefont{and}
  \bibinfo{author}{\bibfnamefont{A.}~\bibnamefont{Vespignani}},
  \bibinfo{journal}{Proceedings of the National Academy of Sciences USA}
  \textbf{\bibinfo{volume}{103}}, \bibinfo{pages}{2015} (\bibinfo{year}{2006}).

\end{thebibliography}

\clearpage

\renewcommand{\theequation}{S\arabic{equation}}
\setcounter{equation}{0}
\renewcommand{\thefigure}{S\arabic{figure}}
\setcounter{figure}{0}
\renewcommand{\thetable}{S\arabic{table}}
\setcounter{table}{0}

\newpage

\section*{Appendix}

\begin{table}
  \begin{center}
    \begin{tabular}{|c|c|c|c|c|}
      \hline
      $\gamma$ & $p_c$ & $\alpha$ & $P_\infty$ & $\beta$
      \\
      \hline \hline
      $2.3$ & $0.18$ &  $0.34$ & $0.01$ & $0.33$
      \\
      \hline
      $2.7$ & $0.25$ & $0.43$ & $0.07$ & $0.39$
      \\
      \hline
      $3.5$ & $0.33$ & $0.60$ & $0.16$ & $0.54$
      \\
      \hline
      $100.0$ & $0.47$ & $0.95$ & $0.32$ & $1.09$
      \\
      \hline
    \end{tabular}
  \end{center}
  \caption{In panel f of Fig.~2,
    we fitted empirical estimations
    of the percolation threshold $p_c(N)$ computed
    at various system sizes $N$ with the function
    $p_c(N) = p_c + N^{-\alpha}$. In Fig.~2g, we perform
    instead the fit
    $P_\infty(N) = P_\infty + N^{-\beta}$
    for the height of the order parameter at criticality.
    Here, we report the values of the
    best estimates of $p_c$, $\alpha$, $P_\infty$
    and $\beta$ for the different
    values of the degree exponent $\gamma$ used in the
    generation of the scale-free networks.}
\end{table}

\subsection*{Taylor expansions}
An alternative way to arrive to the results of
Eq.~(\ref{eq:stability_site_dense})
is to use the multidimensional Taylor expansion
of the r.h.s. of Eq.~(\ref{eq:site_dense}) around
the trivial solution $\vec{s} = \vec{0}$ as
\[
\begin{array}{l}
  [ 1 - \prod_{j \in \mathcal{N}_i} \, (1 - s_j) \, ]
\\
   =
\sum_k \, s_k \,  \left. \frac{d}{d s_k}  \, [ 1 - \prod_{j \in \mathcal{N}_i} \, (1 - s_j) \, ]\right|_{\vec{s} = \vec{0}}  + o(s_i^2)
\\
\simeq  \sum_j A_{ij} \, s_j
\end{array}
 \; .  
 \]
Truncated multidimensional Taylor expansions can
be used also to reduce Eq.~(\ref{eq:site_sparse2})
to Eq.~(\ref{eq:stability_site_sparse}).
The only difference here is that the derivatives
are taken with respect to the variables $r_{i \to j}$, and
the expansion is made around the trivial solution
$\vec{r} = \vec{0}$.

\

When dealing with Eq.~(\ref{eq:interd_dense}),
the Taylor expansion should be instead
extended to at least the second order.
Let us first imagine that the intersection graph
does not contain edges, so that
Eq.~(\ref{eq:interd_dense}) reads as
\[
 s_i = p \;  S_{\mathcal{A-B}_i} \;  S_{\mathcal{B-A}_i} \;.
 \]
Since $S_{\mathcal{A-B}_i}$ and $S_{\mathcal{B-A}_i}$ calculated
at $\vec{s} = \vec{0}$ are zero,
the first derivatives of the r.h.s. calculated  
in $\vec{s} = \vec{0}$ are automatically zero.
The Taylor expansion of r.h.s. is thus  
\[
\begin{array}{l}
  \frac{1}{2} \, \sum_j \sum_k \; \, s_j s_k \; \left. \frac{d^2}{ds_j\, ds_k}  \, S_{\mathcal{A-B}_i} \;  S_{\mathcal{B-A}_i} \right|_{\vec{s}=\vec{0}} + o(s_i^3)
  = 
\\
\frac{1}{2} \, \sum_j \sum_k \; \, s_j s_k \; \left. \frac{d}{ds_j}  \, S_{\mathcal{A-B}_i} \; \frac{d}{ds_k} S_{\mathcal{B-A}_i} \right|_{\vec{s}=\vec{0}} + o(s_i^3)
\end{array}
 \; , 
\]
where the second equality is justified by the fact
that $S_{\mathcal{A-B}_i}$ and $S_{\mathcal{B-A}_i}$ are
zero at $\vec{s} = \vec{0}$.
Using the definitions of $S_{\mathcal{A-B}_i}$ and
$S_{\mathcal{B-A}_i}$ we have that
\[
\left. \frac{d}{ds_j} S_{\mathcal{A-B}_i}\right|_{\vec{s}=\vec{0}} = A_{ij}(1-B_{ij})
\]
and
\[
\left. \frac{d}{ds_j} S_{\mathcal{B-A}_i}\right|_{\vec{s}=\vec{0}} = B_{ij}(1-A_{ij}) \; ,
\]
where $A$ and $B$ are the adjacency matrices of the network layers.
In conclusion, we can approximate
Eq.~(\ref{eq:interd_dense}) in absence of
the intersection term as
\[
s_i =  \frac{p}{2} \, [ \sum_j s_j \, A_{ij}(1-B_{ij}) ] \; [ \sum_j s_j \, B_{ij}(1-A_{ij}) ] \;.
\]
With straightforward considerations, we can also insert
the term accounting for the intersection graph, and write
\[
s_i =  p \, \sum_j A_{ij}B_{ij} s_j
 + \frac{p}{2} \, [ \sum_j s_j \, A_{ij}(1-B_{ij}) ] \; [ \sum_j s_j \, B_{ij}(1-A_{ij}) ]
 \;.
\]
This last equation gives us some insights
about the structure of the solution, but it does not allow
to reduce the original problem to a simple eigenvalue/eigenvector
equation as in the case of isolated networks. Similar considerations
can be deduced by taking the Taylor expansion of
Eq.~(\ref{eq:interd_sparse2}).

\end{document}